\documentclass{aastex}
\usepackage{psfig,emulateapj5}

\shorttitle{Proto-Quasars}
\shortauthors{Kawakatu et al.}

\begin{document}

\title{Proto-Quasars: Physical States and Observable Properties}


\author{Nozomu Kawakatu\altaffilmark{1} and Masayuki Umemura\altaffilmark{2}}
\affil{Center for Computational Physics, University of
  Tsukuba, Tsukuba, Ibaraki, 305-8577, Japan}

\and

\author{Masao Mori\altaffilmark{3}}
\affil{Department of Law, Senshu University, 
Tama-Ku, Kawasaki, 214-8580, Japan}


\altaffiltext{1}{kawakatu@rccp.tsukuba.ac.jp}
\altaffiltext{2}{umemura@rccp.tsukuba.ac.jp}
\altaffiltext{2}{mmori@isc.senshu-u.ac.jp}



\begin{abstract}

  Based on the radiation hydrodynamical model for the black hole (BH) growth, 
incorporated with the chemical evolution of the early-type host galaxy, 
we construct the coevolution model of a QSO BH and the host galaxy.
As a result, it is found that after a galactic wind epoch,
the luminosity is shifted  from the host-dominant phase 
to the AGN-dominant phase (QSO phase) in the timescale of a few $10^{8}$ years.
The former phase corresponds to the early stage of growing BH, and can be 
regarded as a ``proto-QSO'' phase. It has observable characteristic properties 
as follows:
(1) The width of broad emission line is narrower than that of ordinary QSOs,
and it is typically less than 1500km/s.
(2) The BH-to-bulge mass ratio, $M_{\rm BH}/M_{\rm bulge}$, 
is in the range of $10^{-5.3}-10^{-3.9}$.
(3) Host galaxies are bluer compared to QSO hosts,
by about 0.5 magnitude  
in the colors of $({\it B-V})$ at the rest bands  and $({\it V-K})$ 
at the observed bands, 
with assuming galaxy formation redshifts of $z_{\rm f}=3-5$.
(4) The metallicity of gas in galactic nuclei is $\sim 8Z_{\odot}$, and 
that of stars weighted by the host luminosity is $\sim 3Z_{\odot}$.
(5) The central massive BH ($\simeq 10^{7}M_{\odot}$)
is surrounded by a massive dusty disk ($ > 10^{8}M_{\odot}$),
which may obscure the nucleus in the edge-on view
and make a type 2 nucleus.
By comparing these predictions with recent observations, radio galaxies 
are a possible candidate for proto-QSOs.
Also, it is anticipated that the proto-QSO phase is preceded by an optically thick phase, which may correspond to ULIRGs.
In this phase, $M_{\rm BH}/M_{\rm bulge}$ 
is predicted to be much less than $10^{-3}$ 
and grow with metallicity.
Moreover, as precursors of ULIRGs, optically-thin star-forming galaxies 
are predicted. These may be in the assembly phase of Lyman break galaxies (LBGs) 
or Ly$\alpha$ emitters.

\end{abstract}

\keywords{black hole physics---galaxies:active---
galaxies:evolution---galaxies:formation---galaxies: nuclei---
galaxies: starburst---radiation drag}

\section{Introduction}
\label{INTRO}

 Recent X-ray and optical observations suggest the possibility that 
active galactic nuclei (AGNs) could be divided into two subclasses 
according to the rate of black hole (BH) growth; 
one is a rapidly growing phase and the other is a slow growing phase 
(Pounds et al. 1995; Boller et al. 1996; Mineshige et al. 2000; Mathur et al. 2001: 
Wandel 2002).
This possibility has been pointed out primarily 
for the Seyfert 1 galaxies (Sy1s), which are divided 
into two subclasses according to the width of broad emission line, $V_{\rm BLR}$. 
Sy1s with $V_{\rm BLR}$ less than 2000km/s are called narrow line Sy1s (NLSy1s), 
whereas those with broader line width are called broad line Sy1s (BLSy1s).
NLSy1s exhibit two distinctive X-ray properties,
that is, rapid X-ray variability and strong soft X-ray excess. 
These properties can be explained in terms of the optically thick ADAF (advection-dominated 
accretion flow) onto a smaller BH, which is realized by higher accretion rate compared to the 
Eddington limit (Pounds et al. 1995; Boller et al. 1996; Mineshige et al. 2000).
Also, it is pointed out that the BH-to-bulge mass ratio is noticeably smaller than 
that in elliptical galaxies, $M_{\rm BH}/M_{\rm bulge} < 10^{-3}$
(Mathur et al. 2001; Wandel 2002).
All of these suggest that NLSy1s are in the rapidly growing phase of BH, in contrast to BLSy1s which are explained by conventional mild accretion onto a large BH.  
Additionally, Kawaguchi \& Aoki (2000) argue that NLSy1s have a high star formation rate (SFR). 
According to these observations,
it has been suggested that NLSy1s may be Sy1s in the early stage of their 
evolution (Mathur 2000).
On the analogy of NLSy1s, QSOs are also expected to have rapidly growing 
phase of QSO BHs.
But, it has not been elucidated so far what objects correspond 
to the early phase of QSOs.

On the other hand, recent high-resolution observations of galactic centers have revealed that 
the estimated mass of a central ``massive dark object''(MDO), 
which is the nomenclature for a supermassive BH candidate, does correlate 
with the mass of a galactic bulge; 
the mass ratio of the BH to the bulge is 0.001-0.006 as a median value 
(Kormendy \& Richstone 1995; Richstone et al. 1998;
Magorrian et al. 1998; Loar 1998; Gebhardt et al. 2000;
Ferrarese \& Merritt 2000; Merritt \& Ferrarese 2001
McLure \& Dunlop 2001; McLure \& Dunlop 2002;  Wandel 2002).
(It is noted that the bulge means a whole galaxy for an elliptical galaxy.)
In addition, it has been found that QSO host galaxies are 
mostly luminous and well-evolved early-type galaxies 
(McLeod \& Rieke 1995; Bahcall et al. 1997; Hooper, Impey \& Foltz 1997; 
McLoed, Rieke \& Storrie-Lombardi 1999; Brotherton et al. 1999; Kirhakos et al. 1999; 
McLure et al. 1999; McLure, Dunlop \& Kukula 2000).
These findings, combined with the BH-to-bulge relations, suggest that the formation 
of a supermassive BH, an elliptical galaxy, and a QSO is physically related to 
each other.
But, the link between the formation of a supermassive BH and 
the evolution of a host galaxy is an open question.
Also, the physical relationship among QSOs,
ultraluminous infrared galaxies (ULIRGs), and radio galaxies 
has been an issue of long standing.

Some theoretical models of BH growth models have been considered 
to explain the BH-to-bulge correlations 
(Silk \& Rees 1998; Ostriker 2000; Adams, Graff, \& Richstone 2001).
But, little has been elucidated regarding the physics on the angular momentum transfer, 
which is requisite for BH formation.
Recently, as a potential mechanism to remove angular momentum,
Umemura (2001) has considered the effects of radiation drag, 
which is equivalent to a well-known Poynting-Robertson effect.
The exact expressions for the radiation drag are found in the literature 
(Umemura, Fukue, \& Mineshige 1997; Fukue, Umemura, \& Mineshige 1997).
In an optically thick regime, the efficiency of radiation drag is saturated 
due to the conservation of the photon number (Tsuribe, \& Umemura 1997).
Thus, the angular momentum loss rate by the radiation drag is given by 
$d \ln J/dt \simeq -(L_{*}/c^{2}M_{\rm g})$, where $J$, $L_{*}$, 
and $M_{\rm g}$ are 
the total angular momentum of gaseous component, the total luminosity of the bulge, 
and the total mass of gas, respectively.
Then, the maximal rate of mass accretion is given by 
$\dot{M}=-M_{\rm g}d \ln J/dt = L_{*}/c^{2}$ (Umemura 2001).
Thus, the total accreted mass on to the MDO, $M_{\rm MDO}$, is estimated by
\begin{eqnarray}
M_{\rm MDO} &\simeq& \int_{0}^{\infty}\frac{L_{*}}{c^{2}}dt. \nonumber
\end{eqnarray}
In practice, the interstellar medium (ISM) is observed to be highly inhomogeneous 
in an active star-forming galaxies (Sanders et al. 1988; Gordon, Calzetti \& Witt 1997).
Kawakatu \& Umemura (2002) have shown that the inhomogeneity of interstellar 
medium helps the radiation drag to sustain the maximal efficiency.

Thus, the final mass of MDO is proportional to the total radiation energy from bulge stars,
 and the resultant BH-to-bulge mass ratio is basically determined by the energy conversion 
efficiency of the nuclear fusion from hydrogen to helium, i.e. 0.007 (Umemura 2001).
So far, the realistic chemical evolution of the host galaxy has not been
incorporated, but a simple evolutionary model was assumed.
As for the relation between a QSO BH and the host galaxy, 
some phenomenological models have been proposed 
(Haehnelt \& Rees 1993; Haiman \& Loeb 1998; Kauffmann \& Haehnelt 2000; 
Monaco, Salucci, \& Danese 2000; Granato et al. 2001; Hosokawa et al. 2001), 
but little on the physics has been known.
Hence, in order to reveal the formation and evolution of QSOs 
and clarify what objects correspond to the early phase of QSOs,
it is important to investigate the physics on the rapidly growing phase of QSO BHs.
Here, based on the radiation drag model with incorporating the realistic  
chemical evolution, we construct a physical model for the coevolution of a QSO BH and 
the early-type host galaxy.
The purpose of this paper is to elucidate the physical relationship between a BH growth and 
the evolution of host galaxy, and define a proto-QSO phase as an early stage of QSO evolution. 
Then, we predict the observable properties of proto-QSOs.
Also, we address a unified picture for the evolution of an elliptical galaxy nucleus.

The paper is organized as follows.
In section 2, we build up a theoretical model for the coevolution of a QSO BH 
and the early-type host galaxy.
In section 3, we investigate the time-dependent relation between a QSO BH 
and the early-type host galaxy,
and analyze the physical states of proto-QSOs which correspond to 
the rapidly growing phase of a QSO BH.
In Section 4, we propose a unified picture for the evolution of elliptical 
galaxy nucleus.
Section 5 is devoted to the conclusions.

\section{Coevolution model}
\label{R-HM}

First, we model the BH growth based on the radiation drag-driven mass accretion.
Here, we  suppose a two-component system that consists of a spheroidal stellar bulge 
and inhomogeneous optically-thick interstellar medium within it.
The radiation drag efficiency increases with the optical depth $\tau$ in 
proportion to $(1-e^{-\tau})$ (Umemura 2001).
Thus, the mass of an MDO, $M_{\rm MDO}$, 
which is the total mass of dusty ISM assembled to the central massive 
object, is given by 
\begin{equation}
M_{\rm MDO}(t)=\eta_{\rm drag}\int^{t}_{0}\int^{\infty}_{0}
\frac{L_{\rm bulge,\nu}(t)}
{c^2}\left(1-e^{-\tau_{\nu}(t)}\right) d\nu dt,
\end{equation}
where $\tau_{\nu}$ is defined as the optical depth of the bulge measured from the center.
$\eta_{\rm drag}$ is found to be maximally 0.34 in the optically thick regime
(Kawakatu $\&$ Umemura 2002).
Here, we estimate the evolution of $\tau_{\nu}$ for {\it U-, B-, V-} and {\it K-}band
by using an evolutionary spectral synthesis code 'PEGASE' (Fioc \& Rocca-Volmerange 1997).
In Figure 1, the evolution of $\tau_{\nu}$ for all bands is shown. 
We define a time $t_{\rm thin}$ before which the optical depth is less than unity 
in the {\it U}-band, which is most intensive in the early evolutionary phase. 
Before $t_{\rm thin}$, the radiation drag efficiency is low and
the growth of MDO is quite slow.
Furthermore, once a galactic wind occurs, the bulge becomes ISM-deficient and 
thus optically thin, so that the mass accretion via radiation drag is terminated. 
In this paper,
after a wind epoch $t_{\rm w}$ of several $10^{8}$yr, 
$\tau_{\nu}$ is assumed to drop abruptly to $\tau_{\nu}\ll 1$.
Hence, the final mass of an MDO is given by 
\begin{equation}
M_{\rm MDO}=\eta_{\rm drag}\int^{t_{\rm w}}_{t_{\rm thin}}\frac{L_{\rm bulge}(t)}{c^2}dt,
\end{equation}
where $L_{\rm bulge}$ is the bolometric luminosity of the bulge.
It is found that escaping photons from the bulge 
in the optically thin phase of $0<t<t_{\rm thin}$ 
is less than $5\%$ of the total.
%
%
\medskip
\centerline{\psfig{file=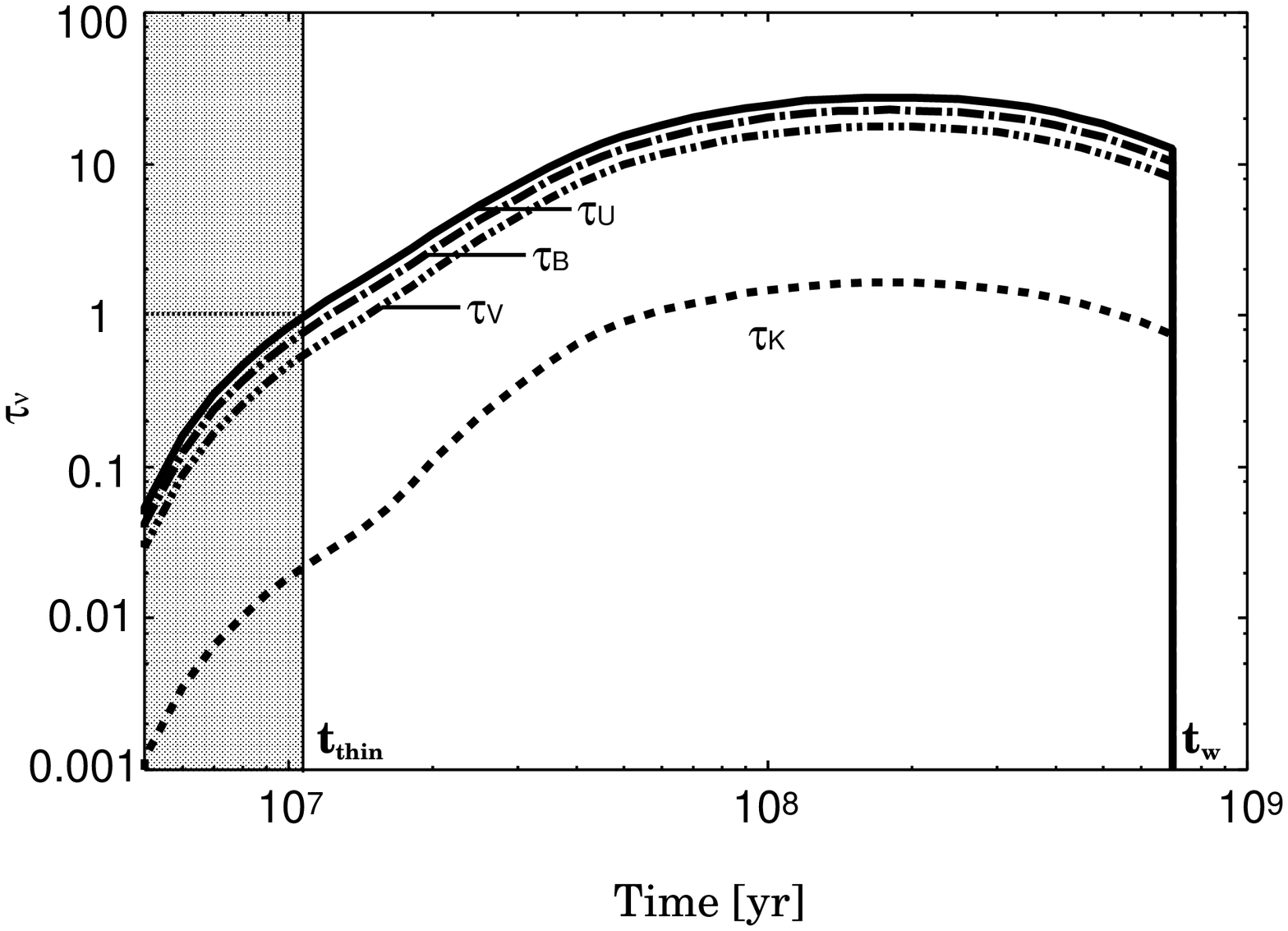,width=3.4in}}
\figcaption[figure1.eps] 
{
The optical depth ($\tau_{\nu}$)  of bulge measured from the center as a function of time.
The abscissa is time in units of yr.
The ordinate is the optical depth for ${\it U-, B-, V-,}$ and ${\it K-}$ band.
A time $t_{\rm thin}$ is defined so that the optical depth is less than unity in the {\it U-}band.
The optically thin phase ($\tau_{U} < 1$) is the gray area.
The system is optically thick until the galactic wind timescale ($t_{\rm w}$), at which  
$\tau_{\nu}$ is assumed to drop abruptly down to $\tau_{\nu} \ll 1$.
\label{Figure 1}
}
\medskip
In this model, we should distinguish the BH mass from the mass of an MDO, 
although the mass of an MDO is often regarded as BH mass from an observational point of view.
Supposing the mass accretion driven by the viscosity on to the BH horizon 
is determined by an order of Eddington rate, 
the BH mass grows according to 
\begin{equation}
M_{\rm BH}=M_{0}e^{\nu t/t_{\rm Edd}},
\end{equation}
 where $\nu$ is the ratio of BH accretion rate to the Eddington rate,  
$\nu=\dot{M}_{\rm BH}/\dot{M}_{\rm Edd}$, 
which is about 0.1 for QSOs (McLeoed, Rieke,\& Storrie-Lombaardi 1999),
and $t_{\rm Edd}$ is the Eddington timescale, 
$t_{\rm Edd}=\eta_{\rm BH}M_{\rm BH}c^{2}/L_{\rm Edd}$, 
where $\eta_{\rm BH}$ is the energy conversion efficiency and 
$L_{\rm Edd}$ is the Eddington luminosity.
Recently, it is shown by a full general relativistic calculation (Shibata \& 
Shapiro 2002) that the collapse of a rotating supermassive star (SMS) results 
in the formation of a rotating BH with Kerr parameter of 0.75.
Hence, $\eta_{\rm drag}$ is assumed to be 0.42, which is the efficiency of 
a extreme Kerr BH.
Then, we have $t_{\rm Edd}=1.9\times 10^{8}$yr.  
$M_{0}$ is the mass of a seed BH, 
which could be an early formed massive BH with $\sim 10^{5}M_{\odot}$ 
(Umemura, Loeb, \& Turner 1993) or a massive BH with $\sim 10^{5}M_{\odot}$ 
formed by the collapse of a rotating supermassive star (SMS)
(Baumgarte \& Shapiro 1999; Saijo et al. 2002; Shibata 2002), 
which may result from by the viscous runaway collapse of the MDO 
(Tsuribe 1999; Umemura 2002).
 
Next, we construct the model for the chemical evolution of host galaxy.
To treat the realistic chemical evolution, 
we use an evolutionary spectral synthesis code 'PEGASE' (Fioc \& Rocca-Volmerange 1997).
In this paper, we consider an elliptical galaxy as a host galaxy to relate the 
formation of a QSO.
Also, we settle the model parameters so that 
the color-magnitude relation of present-day elliptical galaxies can be reproduced.  
In this model, we assume a initial mass function (IMF) as 
$\phi = A(m_{*}/M_{\odot})^{-0.95}$ for a mass range of 
[$0.1,60M_{\odot}$], where $m_{*}$ is the stellar mass. 
The star formation rate (SFR) per unit mass , $C(t), $
is assumed to be proportional to the fractional gas mass 
$f_{\rm g}(t) \equiv M_{\rm g}(t)/M_{\rm g0}$, 
where $M_{\rm g0}$ is the initial gas mass which is $10^{12}M_{\odot}$ 
and $M_{\rm g}(t)$ is the total gas mass at time $t$.
With incorporated a galactic wind model, 
the star formation rate is given by
\begin{eqnarray}
C(t) &=& kf_{\rm g}, \hspace{3cm} (0 \leq t < t_{\rm w}) \nonumber \\
     &=& 0, \hspace{3.42cm} (t \geq t_{\rm w}),       
\end{eqnarray} 
where a constant rate coefficient is set to $k$=8.6Gyr$^{-1}$.
Here, we assume $t_{\rm w}=7\times 10^{8}$ yr from the fiducial wind model
by Arimoto \& Yoshii (1987).
With this chemical evolution model, 
we can pursue the evolution of the physical properties of host galaxy, such as 
stellar mass, luminosity, color, and metallicity.

\section{QSO BH - host relation}
\label{NQF}

Based on the present coevolution model, 
the evolution of the mass of stellar component in the bulge ($M_{\rm bulge}$), 
the mass of MDO ($M_{\rm MDO}$), and the mass of the supermassive BH ($M_{\rm BH}$) 
 are shown in Figure 2, 
assuming the constant Eddington ratio ($\nu =1$).
The mass accretion proportional to the bulge luminosity leads to the growth 
of an MDO up to $10^{8}M_{\odot}$, which is likely to form a massive dusty disk 
in the nucleus.
However, the matter in the MDO does not promptly fall into the BH, 
because the BH accretion is limited by equation(3). 
The BH mass reaches $M_{\rm MDO}$ at a time $t_{\rm cross}$.
As seen in Figure 2, during $t<t_{\rm cross}$ the BH mass fraction 
$f_{\rm BH}=M_{\rm BH}/M_{\rm bulge}$ increases  with time.
At $t>t_{\rm cross}$, almost all of the MDO matter has fallen onto the central BH, 
and therefore the BH mass is saturated.
In the optically thick phase ($t_{\rm thin} < t < t_{\rm w}$), the BH fraction is  
$f_{\rm BH} \ll 0.001$.
In the optically thin phase at $t > t_{\rm w}$, which is the dark gray area
($t_{\rm w} < t < t_{\rm cross}$) in Figure 2, 
the BH fraction increases up to $f_{\rm BH}\simeq 0.001$, 
which is just comparable to the observed ratio.

It has been argued that color-magnitude relation of bulges can be 
reproduced if a galactic wind sweeps away the gas at a wind epoch $t_{\rm w}$ 
of several $10^{8}$yr
(Arimoto \& Yoshii 1986, 1987; Kodama \& Arimoto 1997; Mori et al. 1997).
The evolution of the bulge luminosity is shown in Figure 3 with the 
galactic wind model. 
Even after the galactic wind ($t>t_{\rm w}$), 
$M_{\rm BH}$ continues to grow until $t_{\rm cross}$ 
and therefore the AGN brightens with time if the Eddington ratio is constant.
After the AGN luminosity ($L_{\rm AGN}$) exhibits a peak at $t_{\rm cross}$, 
it fades out abruptly due to exhausting the fuel of the MDO.
As seen in Figure 3, it is found that the era of $t_{\rm w} < t <t_{\rm cross}$ 
can be divided into two phases with a transition time 
$t_{\rm crit}$ when $L_{\rm bulge}=L_{\rm AGN}$; 
the earlier phase is the host luminosity-dominant phase (the dark gray area), 
and the later phase is the AGN luminosity-dominant phase (the light gray area).
The lifetimes of both phases are comparable to each other, 
which is about $10^{8}$yr.
The AGN-dominant phase is likely to correspond to ordinary QSOs, 
but the host-dominant phase is obviously different from observed QSOs so far.
We define this phase as ``a proto-QSO''.
In this phase, $f_{\rm BH}$ rapidly increases
from $10^{-5.3}$ to $10^{-3.9}$ in $\approx 10^{8}$ years.
Also,the central massive BH is surrounded by a massive dusty disk 
($ > 10^{8}M_{\odot}$), which may obscure the nucleus in the edge-on view
and make a type 2 nucleus.
Objects corresponding to such host luminosity-dominant proto-QSOs have not been 
identified observationally yet.
Hence, the detection of this type of QSOs could be a crucial test for the present picture.
The later fading nucleus could be a low luminosity AGN (LLAGN) 
(e.g. Kawaguchi \& Aoki 2001; Awaki et al. 2001). 
The proto-QSO phase is preceded by a bright and optically thick phase,
which may correspond to a ultraluminous infrared galaxy (ULIRG) phase.
\medskip

\centerline{\psfig{file=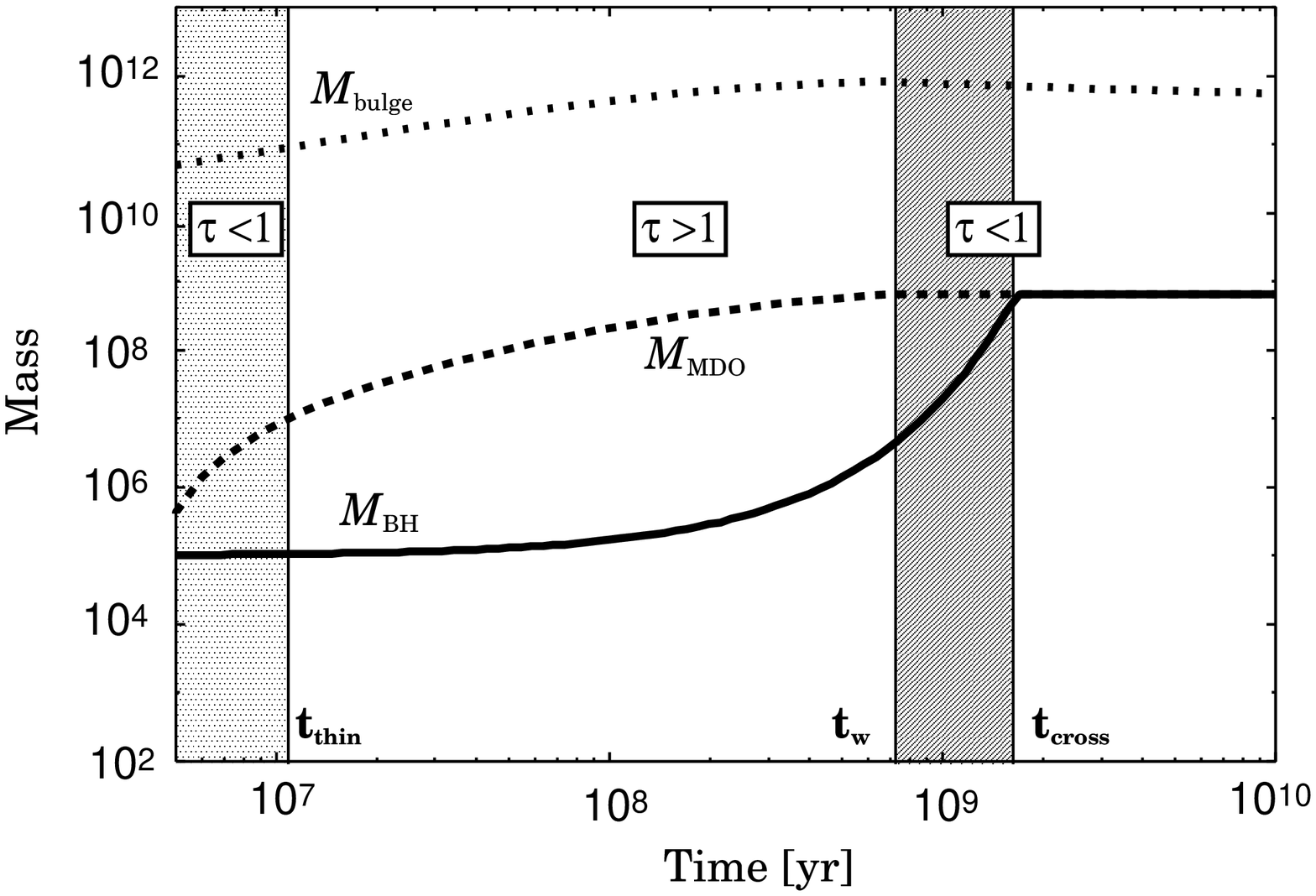,width=3.4in}}
\figcaption[figure2.eps] 
{
BH growth, assuming $M_{0}=10^{5}M_{\odot}$ and $\nu =1.0$
[see equation (3)].
The ordinate is mass in units of $M_{\odot}$.
$M_{\rm bulge}$ is the mass of stellar component in bulge.
$M_{\rm MDO}$ is the mass of MDO.
$M_{\rm BH}$ is the mass of the supermassive BH.
$t_{\rm w}$ is the galactic wind timescale. 
$t_{\rm cross}$ is defined so that $M_{\rm MDO}=M_{\rm BH}$.
The light gray area shows the initial optically thin phase in the 
timescale of $\sim 10^{7}$yr. 
In this phase, there can exist the massive dusty disk ($10^{6-7}M_{\odot}$).
In the optically thick phase ($t_{\rm thin} < t < t_{\rm w}$), the BH fraction 
$f_{\rm BH}=M_{\rm MDO}/M_{\rm bulge} \ll 0.001$.
In the optically thin phase, which is the dark gray area 
($t_{\rm w} < t < t_{\rm cross}$), 
there can be a massive dusty disk ($> 10^{8}M_{\odot}$) around a massive BH, and also  
the BH fraction is $f_{\rm BH}\simeq 0.001$,which is just comparable to the observed ratio.
\label{Figure 2}
}
\medskip
\centerline{\psfig{file=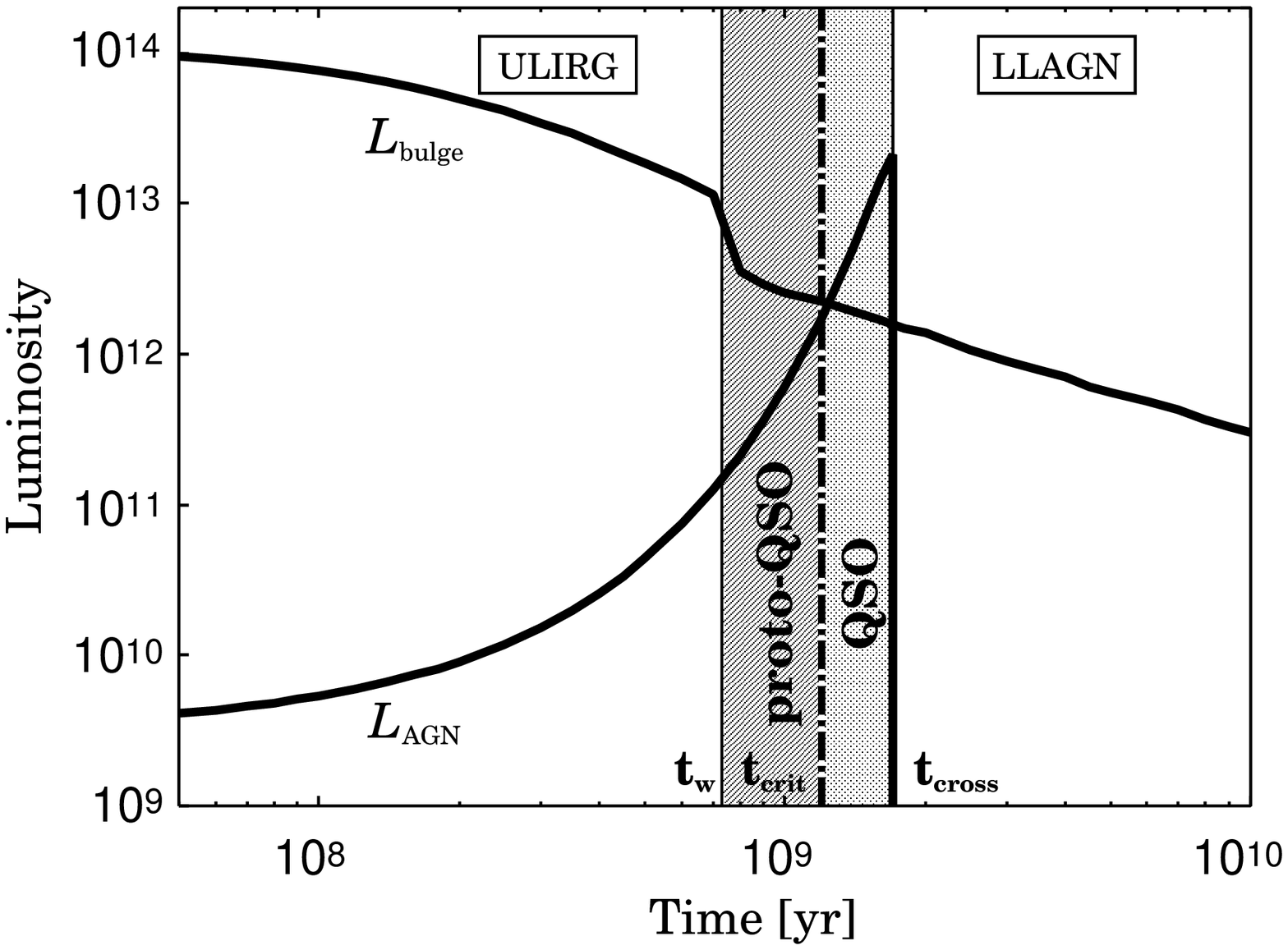,width=3.4in}}
\figcaption[figure3.eps] 
{AGN and bulge luminosity as a function of time.
The ordinate is the luminosity in units of $L_{\odot}$.
$t_{\rm cross}$ is the time when $L_{\rm bulge}=L_{\rm AGN}$.
Here, we assume that $L_{\rm AGN}$ is the Eddington luminosity.
The phase at $t<t_{\rm w}$ is a bright and optically thick phase,
which may correspond to a ultraluminous infrared galaxy (ULIRG) phase.
After the AGN luminosity ($L_{\rm AGN}$) exhibits a peak at $t_{\rm cross}$, 
it fades out abruptly.
The later fading nucleus could be a low luminosity AGN (LLAGN).
The optically-thin, bright AGN phase (gray area) can be divided into two phases; 
one is the host-dominant phase (proto-QSO), which is the dark gray area 
($t_{\rm w}\leq t \leq t_{\rm crit}$) and the other is the AGN-dominant phase
(QSO), 
which is the light gray area ($t_{\rm crit}\leq t \leq t_{\rm cross}$).
The lifetime of both phases are comparable to each other, 
which is about $10^{8}$yr.
}
\label{Figure 3}
\medskip
The proto-QSO phase may be distinguishable in terms of several observable properties like 
the broad emission line width, the color of the host galaxy, and the metallicity.
The time variations of these quantities are shown in Figures 4-6.
The broad emission line width, which corresponds to the virial velocity, $V_{\rm BLR}$, 
is assessed by an empirical law for the size of broad line region (BLR), 
$r_{\rm BLR}=15L_{44}^{1/2}$light-days (Kaspi et al. 1997), 
where $L_{44}$ is the luminosity at $0.1-1\mu {\rm m}$ 
in units of $10^{44}$erg/s.
In Seyfert galaxies, 
this relation holds not only for the slow growing phase of AGN BHs, but also 
for NLSy1s (Peterson et al. 2000).
Provided that the broad line clouds are bound in the potential by central BH (Laor et al. 1997) 
and $L(0.1-1\mu {\rm m})$ is close to the Eddington luminosity, $L_{\rm Edd}\simeq 1.2\times 
10^{46}(M_{\rm BH}/10^{8}M_{\odot})$erg/s,
the circular velocity of broad line clouds ($V_{\rm BLR}$) is given by 
\begin{equation}
V_{\rm BLR}\simeq \left(\frac{GM_{\rm BH}}{r_{\rm BLR}} \right)^{1/2}
=1700\left(\frac{M_{\rm BH}}{10^{8}M_{\odot}}\right)^{1/4}{\rm km/s}.
\end{equation}
The estimated width of broad emission line is shown in Figure 4.
It is found that $V_{\rm BLR}$ in the proto-QSO phase is less than 
$\sim 1500{\rm km/s}$.
This velocity is considerably small compared to ordinary QSOs.
\medskip

\centerline{\psfig{file=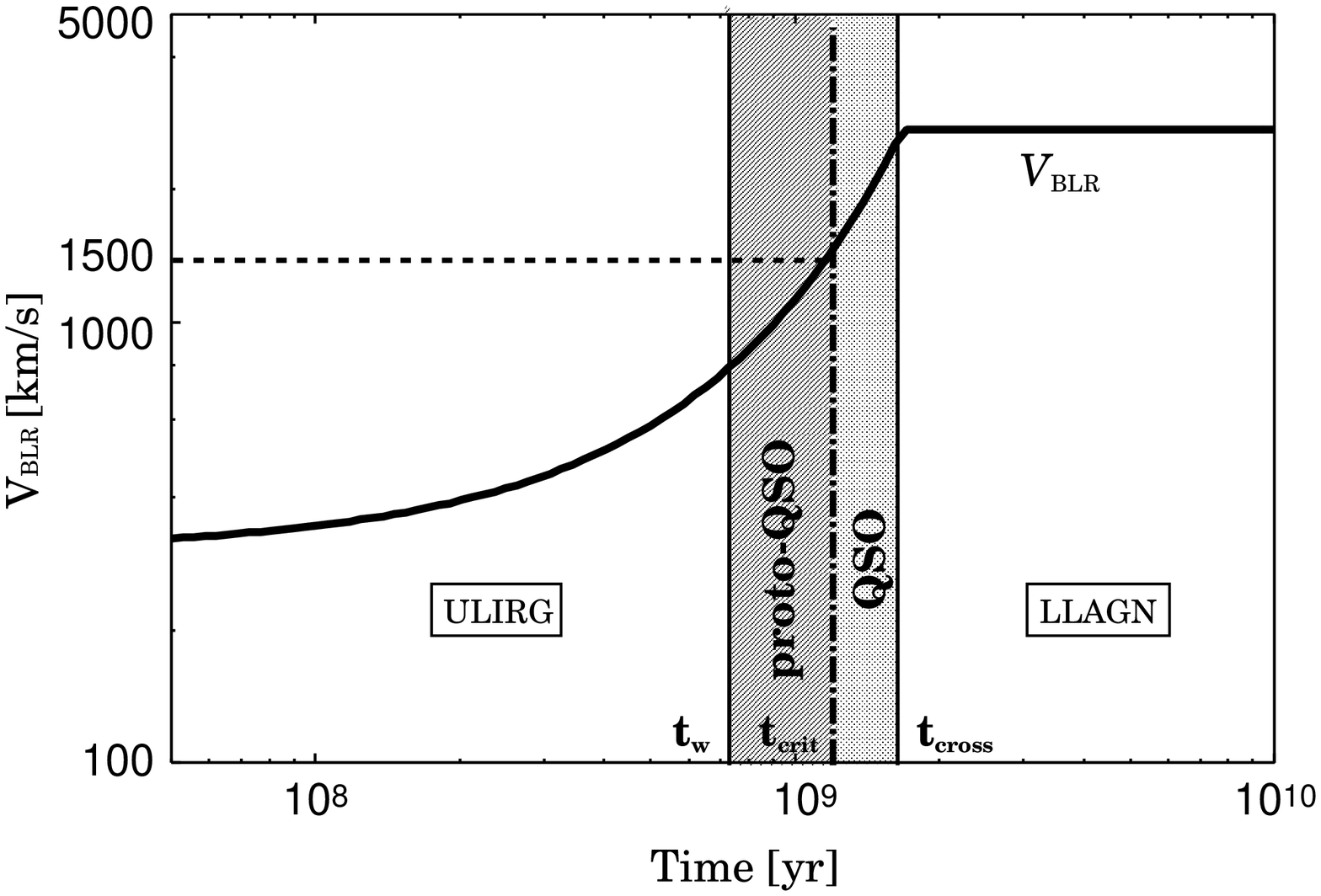,width=3.4in}}
\figcaption[figure4.eps] 
{The broad emission line width, $V_{\rm BLR}$, as a function of time.
The ordinate is $V_{\rm BLR}$ in unit of km/s.
This line represents equation (5).
In the proto-QSO phase, $V_{\rm BLR}$ is less than $\sim$1500km/s.
}
\label{Figure 4}
\medskip
Also, the colors of host galaxy  at the rest and the observed bands 
with assuming the 
formation redshift of $z_{\rm f}=4$ are shown in Figures 5 (a) and (b), respectively.
Here, we take into account the effect of dust extinction and K-correction,
providing  $H_{0}=70$km/s/Mpc and the Einstein-de Sitter universe.
In Figure 5, the thick lines show the cases with dust extinction, 
whereas the thin lines denote the cases without dust extinction.
At the rest frame, as seen in Figure 5 (a), 
it is found that the {\it U-B} color is almost identical between proto-QSO 
and QSO phases,
because the massive stars responsible for {\it U-B} color of 
host galaxy disappear before $\sim 10^{9}$ yr.
However, the {\it B-V} color in the proto-QSO phase 
is about 0.5 magnitude bluer than that in the QSO phase.
Moreover, the {\it V-K} color in the proto-QSO phase 
is about 0.6 magnitude bluer than that in the QSO phase.
At the observed frame, as seen in Figure 5 (b), 
the {\it V-K} color of the proto-QSO phase is about 0.5 magnitude 
bluer than that in the QSO phase, while the noticeable difference is not found in 
{\it U-B} and {\it B-V} colors between two phases. 
We confirm that the results do not change significantly for different redshifts 
of galaxy formation (e.g., $z_{\rm f}$=3 and 5). 
This may be an important photometric feature that can 
discriminate the proto-QSO phase from the QSO phase.

\medskip
\centerline{\psfig{file=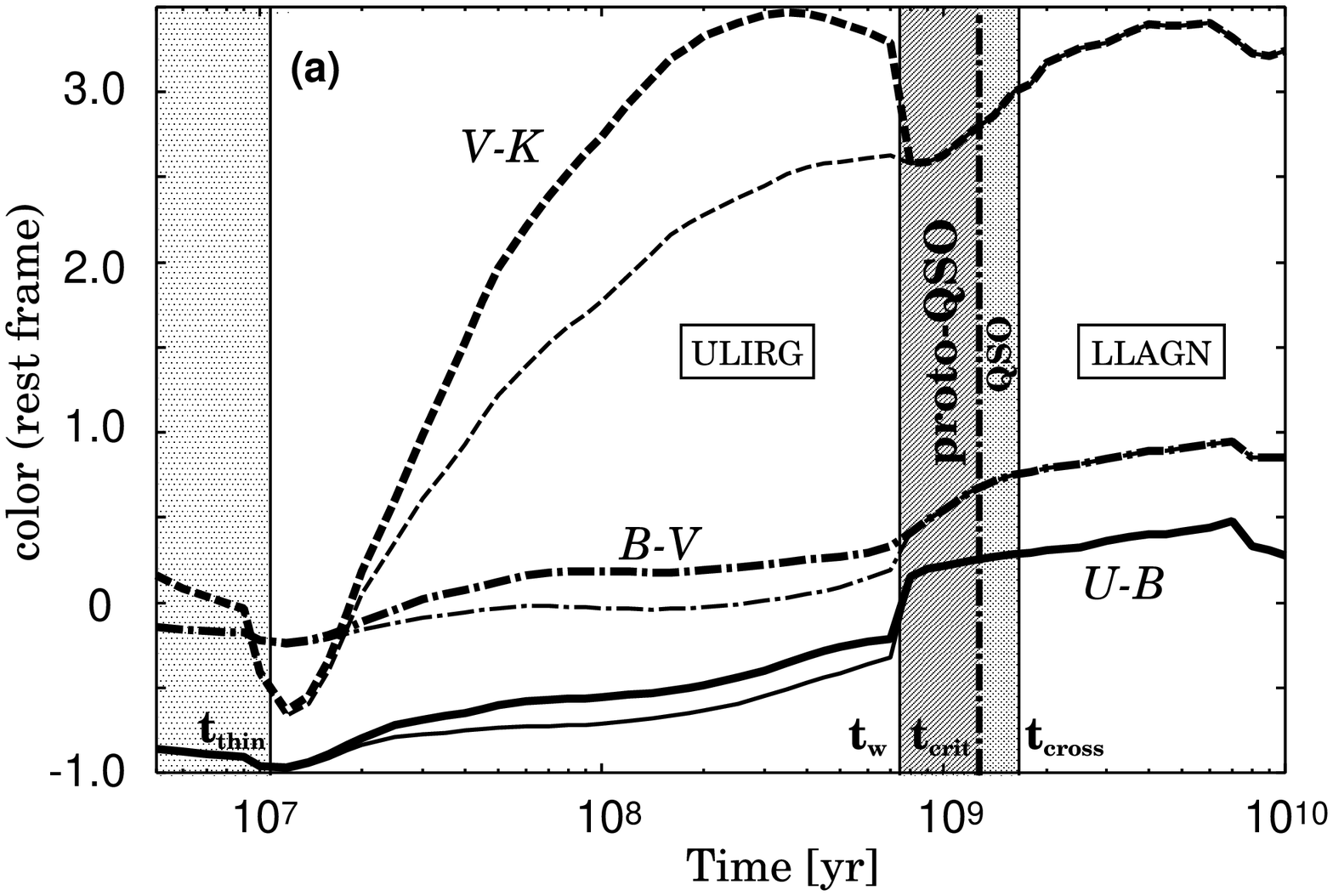,width=3.4in}}
\centerline{\psfig{file=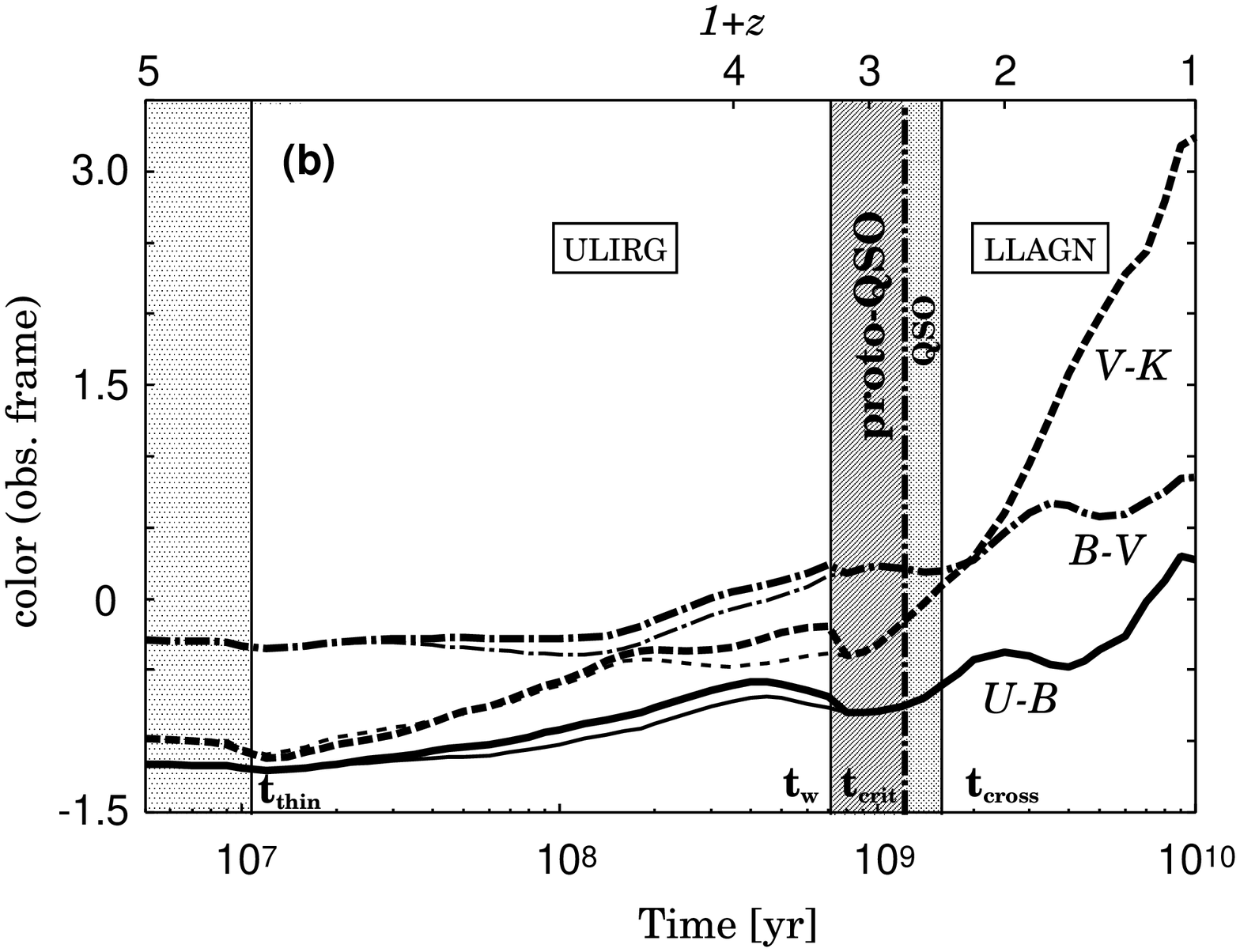,width=3.4in}}
\figcaption[figure5a.eps,figure5b]
{
(a) ${\it U-B, B-V}$ and ${\it V-K}$ colors at the rest frame as a function of time.
The ordinate is the ${\it U-B, B-V}$ and ${\it V-K}$ colors.
The effects of dust extinction and K-correction are taken into account.
The thick lines show the case with dust extinction.
The thin lines denote the case without dust extinction.
The ${\it B-V}$ color in the proto-QSOs is about 0.5 magnitude bluer than in that of QSOs.
Moreover, the ${\it V-K}$ color in the proto-QSO is about 0.6 magnitude bluer than in that of QSOs.   
(b)${\it U-B, B-V}$ and ${\it V-K}$ color at the observed frame as a function of time, 
assuming $H_{0}=70$km/s/Mpc, the galaxy formation redshift of $z_{\rm f}$=4, 
and the Einstein-de Sitter model.
The upper abscissa shows the redshift.
The ${\it V-K}$ color in proto-QSOs is about 0.5 magnitude bluer than the QSO phase.
}
\vspace*{0.3cm}
As for the metallicity, Figure 6 shows that 
no appreciable difference is found between proto-QSO and QSO phases.
The average metallicity of gas in galactic nucleus, 
$Z_{\rm BLR}$ (the solid line), is $\sim 8Z_{\odot}$, and the average metallicity of 
stars weighted by the host luminosity, $Z_{*}$ (the dashed line), is $\simeq 3Z_{\odot}$.
These results are consistent with the observed metallicity in 
QSO BLRs and the inferred metallicity in present-day elliptical galaxies
(Dressler et al. 1984; Hamann \& Ferland 1993; Shields \& Hamann 1997; 
Dietrich \& Whilhelm-Erkens 2000; Hamann, Foltz \& Chaffee 2001; Becker et al. 2001).
\centerline{\psfig{file=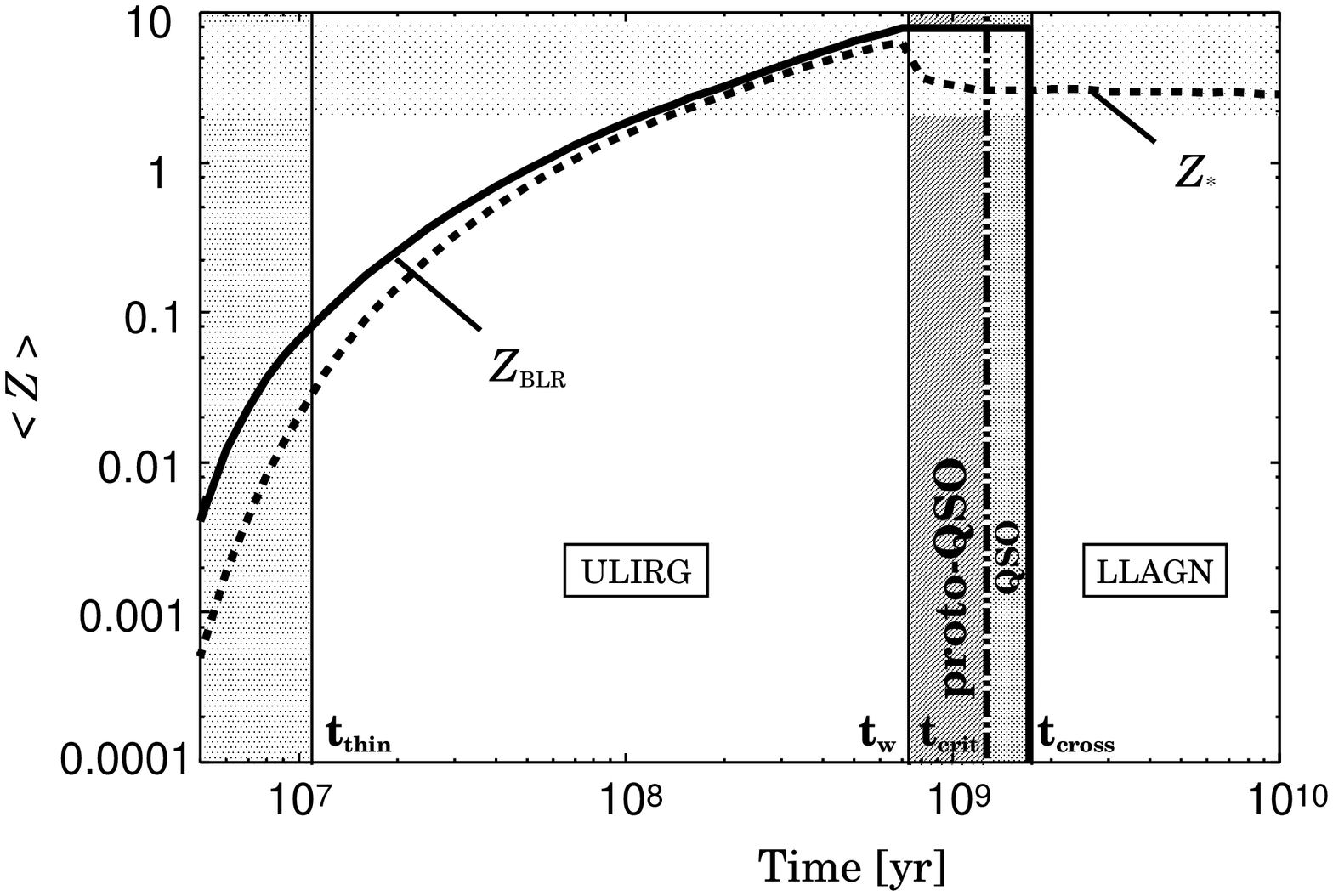,width=3.4in}}
\figcaption[figure6.eps]
{
The mean metallicity $<Z>$ as a function of time.
The ordinate is the metallicity in unit of $Z_{\odot}$.
The solid line denotes the metallicity of gas in galactic nuclei, $Z_{\rm BLR}$.
The dotted line shows the metallicity of stars weighted by the host luminosity, 
$Z_{*}$.
The top hatched area denotes the observable metallicity in QSO BLR 
and inferred metallicity in present-day elliptical galaxies 
(Dressler et al. 1984; Hamann \& Ferland 1993; Shields \& Hamann 1997; 
Dietrich \& Whilhelm-Erkens 2000; Hamann, Foltz \& Chaffee 2001; Becker et al. 2001).
}
%
%
\section{A unified evolutional scenario for an elliptical galaxy nucleus}
Based on the results shown in above sections, we attempt to provide
an evolutionary picture for an elliptical galaxy nucleus.
The present radiation-hydrodynamical model is grounded on two key conditions,
that is, the starburst event and the optically thick interstellar medium.
Obviously, starburst galaxies including ULIRGs satisfy these two.
Starbursts are observed to be often triggered by galaxy mergers or 
interactions (e.g. Borne et al. 2000), and also numerical simulations
of galaxy mergers actually demonstrated the possibility
(e.g., Mihos \& Hernquist 1996). Thus, the present model can be considered
in the context of galaxy mergers.
On the other hand, the recent discovery of high redshift quasars with
$z > 6$ (Fan et al. 2001) indicates that the formation of supermassive BHs 
proceeded in appreciably short timescale less than $10^{9}$yr.
In these cases, there may not be enough time for well-evolved galaxies to merge
to form massive elliptical galaxies.
However, the present model is still valid if QSO hosts satisfy
above two conditions. 

Recent high-quality observations suggest that radio galaxies, ULIRGs, 
Lyman break galaxies (LBGs), and Ly$\alpha$ emitters could be precursors of 
spheroidal galaxies at high redshifts.
As for the radio galaxies,
they are categorized as radio-loud AGNs.
Radio galaxies are thought to have massive BHs ($\sim 10^{8}M_{\odot}$), which is comparable 
to that of the radio-loud QSOs, but the ratio of the AGN luminosity to the 
host luminosity ($L_{\rm AGN}/L_{\rm bulge}$) is less than unity (Dunlop et al. 2001).
In addition, by the analysis of the spectral energy distribution (SED) of 
high redshift ($z > 2$) radio galaxies, the galactic age is estimated to range from $\sim 0.1$ 
to 2Gyr (Mazzei \& Zotti 1996; Pentericci et al. 2001).
Furthermore, a large amount of dust is detected (e.g., Papadopoulos et al. 
2000), and the gas extended over several tens of kpc 
has solar or supersolar metallicity (Vernet et al. 2001, Villar-Martin et al. 2001).
It is intriguing that all these properties of radio galaxies are quite similar to the 
predicted properties of proto-QSOs.
In that sense, radio galaxies could be a key candidate for proto-QSOs.

As for ULIRGs, the X-ray emission (Brandt et al. 1997) or 
Pa$\alpha$ lines (Velleux, Sanders \& Kim 1999) intrinsic for active nuclei have been detected in 
roughly one forth of ULIRGs. Also,
recent X-ray observations have revealed that ULIRGs with AGN activity mostly 
have the lower ratio 
of the hard X-ray luminosity to the bolometric luminosity, which is 
$L_{\rm x}/L_{\rm bol} \ll 0.01$ (Imanishi et al. 1999; Braito et al. 2002).
If the AGN luminosity is controlled by the Eddington limit, 
these results indicate that the BH mass is considerably small in a ULIRG phase. 
In the present model, if the optically-thick phase ($t_{\rm thin} < t < t_{\rm w}$) is 
regarded as a ULIRG phase,
the mass ratio $M_{\rm BH}/M_{\rm bulge}$ is predicted to be much  
less than 0.001 in the ULIRG phase.
Also, it is expected that $M_{\rm BH}/M_{\rm bulge}$ grows with metallicity at the later phases 
of ULIRGs.
The present model may be a physical picture of the evolution of ULIRGs to QSOs 
proposed by Sanders et al. (1988) and Norman \& Scovill (1988). 

The present picture also predicts the existence of the optically thin phase 
($t<t_{\rm thin}$) before the ULIRG phase as shown in Figure 1.
This phase has several characteristic properties;
i) The hard X-ray luminosity is relatively low $L_{\rm x}=5\times 10^{8}L_{\odot}$ 
if $L_{\rm x}=0.1L_{\rm AGN}$.
ii) The metallicity of gaseous component is subsolar, $Z_{\rm BLR} < 0.1Z_{\odot}$ (Figure 6).
iii) The massive dusty disk with $10^{6-7}M_{\odot}$ can surround the galactic center (Figure 2). 
iv) The seed BH with $\sim 10^{5}M_{\odot}$ can form due to the collapse of 
a rotating supermassive star (SMS).
Then, the gravitational wave is expected to be emitted, 
which may be detectable by the Laser Interferometer Space 
Antenna (LISA) (de Araujo et al. 2001; Saijo et al. 2002).
As for the host galaxy, recently observed high redshift LBGs or 
Ly$\alpha$ emitters may correspond to the beginning phase of a bulge
(Friaca \& Terlevich 1999; Matteucci \& Pipino 2002). 
LBGs have the luminosity $10^{10-11}L_{\odot}$, and observed to be optically thin and have the metallicity of $0.1-1Z_{\odot}$.
Also, they exhibit strong clustering at $z\sim 3$.
In addition, the chandra X-ray observatory has detected the hard X-ray
of LBGs with the luminosity of $\sim 10^{8}L_{\odot}$, 
although it is still uncertain whether the X-ray emission arises from AGN or not (Brandt et al. 2001).
Some LBGs have a light profile following a $r^{1/4}$ law over a large radial range 
(Giavalisco et al. 1996).
Moreover, it has been proposed that Ly$\alpha$ emitters may be the first $10^{7}$ yr of the galaxy 
formation, and thereafter their luminosity fades by dust rapidly (Malhotra et al. 2002).
Therefore, the precursor of ULIRGs predicted in the present model may correspond to 
the assembly phase of LBGs or Ly$\alpha$ emitters. 

\section{Conclusions}
\label{D}
Based on the radiation drag model for the BH growth, incorporated with the chemical evolution of 
the early-type host galaxy, we have built up the coevolution model for a QSO BH and the host galaxy.
As a consequence, we have shown the possibility of the proto-QSO phase, 
which is optically-thin and host luminosity-dominant, and has the life-time 
comparable to the QSO phase timescale of a few $10^{8}$ years.
We have predicted the observable properties of proto-QSOs as follows:
(1) The width of broad emission line is narrower, which is less than $1500$km/s.
(2) The BH-to-bulge mass ratio, $M_{\rm BH}/M_{\rm bulge}$, rapidly increases
from $10^{-5.3}$ to $10^{-3.9}$ in $\approx 10^{8}$ years.
(3) The colors of $({\it B-V})$ at rest bands and $({\it V-K})$ at observed bands 
are about 0.5 magnitude bluer than those of QSOs.
(4) In both proto-QSO and QSO phases, the metallicity of gas in galactic nuclei is 
$Z_{\rm BLR}\simeq 8Z_{\odot}$, 
and that of stars weighted by the host luminosity is $Z_{*}\simeq 3Z_{\odot}$,
which are consistent with the observations for QSOs and the elliptical galaxies.
(5) A massive dusty disk ($ > 10^{8}M_{\odot}$) surrounds a massive BH, 
and it may obscure the nucleus in the edge-on view to form a type 2 nucleus.
The predicted properties of proto-QSOs are similar to those of radio galaxies.

The proto-QSO phase is preceded by an optically-thick phase before the galactic wind, 
which may correspond to ULIRGs.
The present model predicts a low luminosity ratio of $L_{\rm AGN}$ to $L_{\rm bol}$, 
which is consistent with the observed ratio $L_{\rm x}/L_{\rm bol} \ll 0.01$ for ULIRGs.
In addition, $M_{\rm BH}/M_{\rm bulge}$ is anticipated to be much less than $10^{-3}$ and 
 $M_{\rm BH}/M_{\rm bulge}$ grows with metallicity in the ULIRG phase.

Finally, we can predict the precursor of ULIRGs, which is optically thin and their lifetime is 
$\sim 10^{7}$ years. 
This may correspond to the assembly phase of LBGs or Ly$\alpha$ emitters.
In this phase, the massive dusty disk of $\approx 10^{6-7}M_{\odot}$ exists, 
the metallicity is subsolar ($Z_{*} < 0.1Z_{\odot} $), 
and the hard X-ray luminosity is $L_{\rm x}\sim 5\times 10^{8}L_{\odot}$ 
if $L_{\rm x}=0.1L_{\rm AGN}$.
In addition, the formation of a seed black hole (BH) ($\sim 10^{5}M_{\odot}$) can  
occur due to the collapse of a rotating supermassive star (SMS) in this phase.
Thus, the gravitational wave may be detectable by the Laser Interferometer Space Antenna (LISA).  
 
\acknowledgments
We thank T.Nakamoto and H.Susa for frutiful discussions.
We are grateful to K.Ohsuga and A.Yonehara for many useful comments.
We also thank the anonymous referee for valuable comments.
Numerical simulations were performed with facilities at the Center of Computational Physics, 
University of Tsukuba. This work was supported in part by the Grant-in-Aid of the JSPS, 11640225.



\begin{thebibliography}{99}
\bibitem[]{}
Adams, F., Graff, D.S., \& Richstone, D.O. 2001, ApJ, 551, L31
\bibitem[]{}
Arimoto, N., \& Yoshii, Y. 1986, A\&A, 164, 260
\bibitem[]{} 
Arimoto, N., \& Yoshii, Y. 1987, A\&A, 173, 23
\bibitem[]{}
Awaki, H., Terashima, Y., Hayashida, K., \& Sakano, M. 2001, PASJ. 53, 647
\bibitem[]{}
Bahcall, J.N., et al. 1997, ApJ, 479, 642
\bibitem[]{}
Baumgarte, T.W., \& Shapiro, S.L. 1999, ApJ, 526, 941 
\bibitem[]{}
Becker, R.H., et al. 2001, AJ, 122, 285 
\bibitem[]{}
Braito, V., et al. 2002, submitted (astro-ph/0202352)
\bibitem[]{}
Brandt, W.N., et al. 2001, ApJ, 558, L5
\bibitem[]{}
Brandt, W.N., et al. 1997, MNRAS, 290, 617
\bibitem[]{}
Brotherton, M.S., et al. 1999, ApJ, 520, L87
\bibitem[]{}
Boller, Th., Brandt, W.N., \& Fink, H. 1996, A\&A, 305, 53
\bibitem[]{}
Borne, K. D., Bushouse, H., Lucas, R. A., \& Colina, L. 2000, ApJ, 529, L77
\bibitem[]{}
De Arajo, J.C.N., Miranda, O.D., \& Agular, O.D. 2001, ApJ, 550, 368
\bibitem[]{}
Dietrich, M., \& Wilhelm-Erkens, U. 2000, A\&A, 354, 17
\bibitem[]{}
Dressler, A.A. 1984, ApJ, 281, 512
\bibitem[]{}
Dunlop, J.S., et al. 2001, MNRAS, submitted (astro-ph/0108397)
\bibitem[]{}
Fan, X. et al. 2001, AJ, 122, 2833
\bibitem[]{}
Ferrarese, L., \& Merritt, D. 2000, ApJ, 539, L9
\bibitem[]{}
Fioc, M., \& Rocca-Volmerrange, B. 1997, A\&A, 326, 950
\bibitem[]{}
Friaca, A.C.S., \& Terlevich, R.J. 1999, MNRAS, 305, 90
\bibitem[]{}
Fukue, J., Umemura, M., \& Mineshige, S., 1997, PASJ, 49, 673
\bibitem[]{}
Gebhardt, K., et al. 2000, ApJ, 543, L5
\bibitem[]{}
Giavalisco, M., Steidel, C.C., Macchetto, F.D., 1996, ApJ, 470, 189 
\bibitem[]{}
Gordon, K., Calzetti, D., \& Witt, A.N, 1997, ApJ, 487, 625
\bibitem[]{}
Granato, G.L., Silva, L., Monaco, P., Panuzzo, P., Salucci, P., 
De Zotti, G., \& Danse, L. 2001, MNRAS, 324, 757
\bibitem[]{}
Haehnelt, M.G., \& Rees, M.J. 1993, MNRAS, 263, 168
\bibitem[]{}
Hamann, F., \& Ferland, G. 1993, ApJ, 418, 11
\bibitem[]{}
Hamann, F., Foltz, C.B., \& Chaffee, F.H. 2001, preprint (astro-ph/0109328) 
\bibitem[]{}
Haiman, Z., \& Loeb, A. 1998, ApJ, 503, 505
\bibitem[]{}
Hooper, E.J., Impey, C.D., \& Foltz, C.B. 1997, ApJ, 480, L95
\bibitem[]{}
Hosokawa, T., Mineshige, S., Kawaguchi, T., Yoshikawa, K., \& Umemura, M. 
2001, PASJ, 53, 861
\bibitem[]{}
Imanishi, M., \& Ueno, M. 1999, ApJ, 527, 709
\bibitem[]{}
Kaspi, S. 1997, in Emission Lines in Active Galaxies: 
New Methods and Techniques, ed.B.M.Peterson, F.-Z.Cheng, \& A.S.Wilson 
(San Fransisco: ASp), 159
\bibitem[]{}
Kauffmann, G., \& Haehnelt, M. 2000, MNRAS, 311, 576
\bibitem[]{}
Kawaguchi, T., \& Aoki, K. 2001, PASJ, submitted
\bibitem[]{}
Kawaguchi, T., \& Aoki, K. 2000, submitted
\bibitem[]{}
Kawakatu, N., \& Umemura, M. 2002, MNRAS, 329,572
\bibitem[]{}
Kirhakos, S., Bahcall, J.N., Schneider, D.P., \& Kristian, J. 1999, 
ApJ, 520, 67
\bibitem[]{}
Kodama, T., \& Arimoto, N. 1997, A\&A, 320, 41
\bibitem[]{}
Kormendy, J., \& Richstone, D. 1995, ARA\&A, 33, 581
\bibitem[]{}
Laor, A. 1998, ApJ, 505, L83
\bibitem[]{}
Laor, A., Fiore, F., Elvis, M., Wilkes, B.J., McDowell, J.C. 1997, 
ApJ, 477, 93
\bibitem[]{}
Magorrian, J., et al. 1998, AJ, 115, 2285
\bibitem[]{}
Malhotra, S, \& Rhoads, J.E. 2002, 565, L71
\bibitem[]{}
Mathur, S., Kurazkiewicz, J. \& Czerny, B.  2001, New Astronomy, 6, 321
\bibitem[]{}
---------. 2000, MNRAS, 314, L17
\bibitem[]{}
Matteucci, F., \& Pipino, A. 2002, ApJ, 569, L69
\bibitem[]{}
Mazzei, P., \& Zotti, G.D., 1996, MNRAS, 279, 535
\bibitem[]{}
McLeod, K.K., \& Rieke, G.H. 1995, ApJ, 454, L77
\bibitem[]{}
McLeod, K.K., Rieke, G.H., \& Storrie-Lombardi, L.J. 1999, ApJ, 511, L67
\bibitem[]{}
McLure, R.J., \& Dunlop, J.S. 2002, MNRAS, 331, 795
\bibitem[]{}
-----------. 2001, MNRAS, 327, 199 
\bibitem[]{}
McLure, R.J., Dunlop, J.S., \& Kukula, M.J. 2000, MNRAS, 318, 693
\bibitem[]{}
McLure, R.J., Kukula, M.J., Dunlop, J.S., Baum, S.A., O'Dea, C.P., \& 
Hughes, D.H. 1999, MNRAS, 308, 377
\bibitem[]{}
Merritt, D., \& Ferrarese L. 2001, MNRAS, 320, L30
\bibitem[]{}
Mihos, J. C., \& Hernquist, L. 1996, ApJ, 464, 641
\bibitem[]{}
Mineshige, S., Kawaguchi, T., Takeuchi, M., \& Hayashida, K. 2000, PASJ, 
52, 499
\bibitem[]{}
Monaco, P., Salucci, P., \& Danese, L. 2000, MNRAS, 311, 279
\bibitem[]{}
Mori, M., Yoshii, Y., Tsujimoto, T., Nomoto, K. 1997, ApJ, 478, L21
\bibitem[]{}
Norman, C., \& Scoville, N. 1988, ApJ, 332, 124
\bibitem[]{}
Ostriker, J.P. 2000, Phys.Rev.Lett., 84, 5258
\bibitem[]{}
Pentericci, L., et al. 2001, ApJS, 135, 63
\bibitem[]{}
Peterson, B.M., McHardy, I.M., \& wilkes, B.J. 2000, New Astronomy Reviews, 
44, 491
\bibitem[]{}
Papadopoulos, P.P., et al. 2000, ApJ, 528, 626
\bibitem[]{} 
Pounds, K.A., Done, C., \& Osbore, J.P. 1995, MNRAS, 277, L5
\bibitem[]{}
Richstone, D.,  et al. 1998, Nature, 395A, 14
\bibitem[]{}
Saijo, M., Baumgarte, T.W., Shapiro, S.L., \& Shibata M. 2002, ApJ, 569,349
\bibitem[]{} 
Sanders D.B., et al. 1988, ApJ, 325, 74
\bibitem[]{}
Shibata, M., \& Shapiro, S.,  2002, ApJ, 572, L39
\bibitem[]{}
Shields, J.C., \& Hamann, F. 1997, RevMexAA, 6, 221
\bibitem[]{}
Silk, J., \& Rees, M.J. 1998, A\&A, 331, L1
\bibitem[]{}
Tsuribe, T., \& Umemura, M., 1997, ApJ, 486, 48
\bibitem[]{}
Tsuribe, T., 1999, ApJ, 527, 102
\bibitem[]{}
Veilleux, S., Sanders, D.B., \& Kim, D.-C. 1999, ApJ, 522, 139
\bibitem[]{}
Vernet, J., et al., 2001, A\&A, 366, 7
\bibitem[]{}
Villar-Martin, M., et al. 2001, ApSSS, 277, 571
\bibitem[]{}
Umemura, M., 2002, The 11th Workshop on General Relativity and Gravitation, eds. K.Maeda et al., 
p.48
\bibitem[]{}
Umemura, M. 2001, ApJ, 560, L29
\bibitem[]{}
Umemura, M., Fukue, J., \& Mineshige, S., 1997, ApJ, 479, L97
\bibitem[]{}
Umemura, M., Loeb, A., \& Turner, E.L. 1993, ApJ, 419, 459
\bibitem[]{} 
Wandel, A. 2002, AJ, 565, 762
\end{thebibliography}
\end{document}